\documentclass[12pt]{article}
\usepackage{amsmath}
\usepackage{amsmath,amssymb}
\usepackage[left=2.5cm,top=2cm,right=2.5cm]{geometry} 
\usepackage{slashed}
\newcommand{\be}[1]{ \begin{equation}\label{#1} }
\newcommand{\ee}{\end{equation}}
\newcommand{\bea}[1]{\begin{eqnarray}\label{#1} }
\newcommand{\eea}{\end{eqnarray}}

\def\ZZZ{{\hskip-3pt\hbox{ Z\kern-1.6mm Z}}}
\def\zzz{{\hskip-3pt\hbox{ z\kern-1mm z}}}
\usepackage{amstext,amssymb,amsfonts}

\newcommand{\ads}{{\text{AdS}_5}}

\newcommand{\ben}{\begin{eqnarray}\displaystyle}
\newcommand{\een}{\end{eqnarray}}

\def\one{{\hbox{ 1\kern-.8mm l}}}
\def\zero{{\hbox{ 0\kern-1.5mm 0}}}

\def\l{\left}
\def\r{\right}

\def\s{\sigma}

\def\o[#1]{{\rm O}\left({#1}\right)}
\def\dotl[#1,#2]{\left\langle #1, #2 \right\rangle}
\def\dotlb[#1,#2]{[ #1, #2 ]}
\def\dotp[#1,#2]{(#1) \cdot (#2)}
\def\>{\rangle}
\def\<{\langle}

\def\ss2{\text{S}^2}

\def\adss{\text{AdS}_2}

\def\s2s2{\ss2\otimes\ss2}
\def\ads2s2{\adss\otimes\ss2}
\def\s2s2n{\l(\ss2\otimes\ss2\r)/\mathbb{Z}_N}
\def\ads2s2n{\l(\adss\otimes\ss2\r)/\mathbb{Z}_N}

\DeclareMathOperator{\expo}{exp}

\begin{document}
\baselineskip 24pt

\begin{center}
{\Large \bf Logarithmic Corrections to Entropy of Magnetically Charged AdS$_4$ Black Holes }

\end{center}

\vskip .6cm
\medskip

\vspace*{4.0ex}

\baselineskip=18pt

\centerline{\large \rm Imtak Jeon$^a$, Shailesh Lal$^b$}

\vspace*{4.0ex}

\centerline{\large \it $^a$Harish-Chandra Research Institute}
\centerline{\large \it  Chhatnag Road, Jhusi,
Allahabad 211019, India}
\centerline{\large \it $^b$LPTHE -- UMR 7589, UPMC Paris 06,}
\centerline{\large \it  Sorbonne Universit{\'e}s,  Paris 75005, France}

\vspace*{1.0ex}
\centerline{\small E-mail:  imtakjeon@gmail.com, shailesh@lpthe.jussieu.fr}

\vspace*{5.0ex}

\centerline{\bf Abstract} \bigskip

We compute logarithmic corrections to the entropy of a magnetically charged extremal black hole in AdS$_4 \times S^7$ using the quantum entropy function and discuss the possibility of matching against recently derived microscopic expressions.

\vfill \eject

\baselineskip=18pt
\tableofcontents
\section{Introduction}
Providing a microscopic interpretation to the Bekenstein Hawking formula in the context of certain classes of supersymmetric extremal black holes in flat space has been a main success of string theory as a theory of quantum gravity \cite{Strominger:1996sh,Dijkgraaf:1996it,Pioline:2005vi,Shih:2005qf,David:2006yn,Sen:2008ta}. The expression for the microscopic entropy obtained by explicit enumeration and counting of black hole microstates in these cases contains the area law as the leading formula, but also contains higher-derivative and quantum corrections to it. We refer the reader to \cite{Sen:2007qy} for a review of these developments as well as more exhaustive references. Importantly, since extremal black holes expectedly possess an AdS$_2$ factor in the near horizon geometry, one may use the AdS$_2$/CFT$_1$ correspondence to provide an alternative, but equivalent, definition of the quantum entropy of extremal black holes in string theory. This proposal is known as the quantum entropy function, and 
for extremal black holes carrying charges $\vec{q}\equiv q_i$ \cite{Sen:2008yk,Sen:2008vm},
\begin{equation}\label{qef}
d_{hor}\l(\vec{q}\r)\equiv\l\langle\exp\l[i\oint q_i d\theta\mathcal{A}_\theta^i\r] \r\rangle_{\adss}^{finite},
\end{equation}
where $d_{hor}$ is the full quantum degeneracy associated with the black hole horizon, and $\mathcal{A}_\theta^i$ is the component of the $i^{\text{th}}$ gauge field along the boundary of the AdS$_2$. In this picture the entropy associated to the horizon degrees of freedom of an extremal blackhole is essentially the free energy corresponding to the partition function \eqref{qef}. The superscript `finite' reminds us that the quantity on the right hand side of \eqref{qef} is naively divergent due to the infinite volue of AdS$_2$ but this divergence may be regulated in accordance with general principles of the AdS/CFT correspondence and a cutoff-insensitive finite part extracted, which is then identified to $d_{hor}$ \cite{Sen:2008yk,Sen:2008vm,Sen:2009vz}. The path integral is carried out over all fields that asymptote to the black hole near horizon geometry. In the context of supersymmetric extremal black holes in flat space, this has been evaluated using saddle point techniques \cite{Sen:2009vz,Banerjee:2010qc,Banerjee:2011jp,Sen:2011ba,bmpvlog,Gupta:2013sva,Gupta:2014hxa} as well as supersymmetric localization \cite{Dabholkar:2010uh,Dabholkar:2011ec,Gupta:2012cy,Dabholkar:2014ema,Murthy:2015yfa,Gupta:2015gga,Murthy:2015zzy} and the answer matched with microscopic results wherever available. Importantly, even in the cases where the full microscopic formula is unavailable, this quantity may be evaluated at least using saddle-point methods to gain some insight into the full microscopic formula \cite{Sen:2011ba,Gupta:2014hxa}.

In contrast, the situation for entropy computations extremal black holes in AdS space is relatively in its infancy. While the Wald entropy may be computed for such black holes using for example the entropy function formalism, the quantum corrections to it are still unknown. In this situation, it would clearly be of interest to evaluate \eqref{qef} for such black holes to obtain the set of quantum corrections to the Wald formula. Such computations are further motivated by the recent proposal for a \textit{microscopic} computation of a CFT$_3$ index argued to capture the quantum entropy of an AdS$_4$ extremal black hole \cite{Benini:2015eyy}.

In this note we will focus on the computation of \eqref{qef} in the semi-classical approximation where the black hole has a large length scale $a$ associated with it, which sets the scale for the area $A_H$ of the event horizon. In the case of extremal black holes in flat space, this large length scale arises when the charges of the black hole are taken to be large. In the present case of AdS black holes, this corresponds to taking the the rank $N$ of the gauge group in the dual CFT to be large, while the charges themselves are not scaled.

One saddle--point of the path integral \eqref{qef} is the black hole's near--horizon geometry itself. By evaluating the on--shell action on this field configuration, it is possible to show that \cite{Sen:2008yk,Sen:2008vm}
\begin{equation}
d_{hor}\simeq e^{S_{\text{Wald}}},
\end{equation}
and hence the quantum entropy function produces as the leading contribution, Wald's formula for black hole entropy \cite{Wald:1993nt}. Now, the next term in the large--charge expansion of the black hole entropy is the so called \textit{log term}, i.e.\footnote{Here we take the leading answer to be Bekenstein--Hawking.}
\begin{equation}
S_{\text{quantum}}= {A_{H}\over 4} + c\ln A_{H},
\end{equation}
where $c$ is a coefficient which depends on the details of the quantum gravity that the black hole is embedded in. For example, the same four--dimensional black hole which is a quarter--BPS black hole in $\mathcal{N}=4$ supergravity may be embedded as a one--eighth BPS black hole in $\mathcal{N}=8$ supergravity and while the leading Bekenstein--Hawking answer is the same for the black hole, the log terms computed in both theories are different \cite{Banerjee:2011jp}, and match with the microscopic computations carried out respectively in $\mathcal{N}=4$ and $\mathcal{N}=8$ string compactifications. This matching is an important test of the consistency of the quantum entropy function proposal.

The main reason why the log term is an important contribution to the microscopic formula is that it is a genuinely quantum correction to the Bekenstein Hawking formula, but is determined completely from one-loop fluctuations of massless fields of two-derivate supergravity, which essentially constitute the IR data of the black hole \footnote{The term `massless' has to be carefully defined on curved manifolds. The more precise statement is that the eigenvalues of the kinetic operator should scale as ${1\over a^2}$, which is how the eigenvalues of the kinetic operator over a massless scalar, $-\Delta$ would scale.}. To see this, let us recall some elements of a scaling argument presented in \cite{Sen:2012dw}. Consider the $\ell$-loop free energy for a theory defined on a $D$ dimensional background with a length scale $a$ associated with it. A typical Feynman graph contributing to this quantity would scale as (see \cite{Sen:2012dw} for details)
\begin{equation}\label{scaling}
\ell_P^{(D-2)(\ell-1)}a^{-(D-2)(\ell-1)}\int^{a/\sqrt{\epsilon}} d^{D\ell}\tilde{k} \,\tilde{k}^{2-2\ell} F\left(\tilde{k}\right),
\end{equation}
where $\tilde{k}=ka$ and $\lbrace\,k\,\rbrace$ are the loop momenta, and $F$ is a function which approaches 1 at large values of its arguments. By focusing on the regime where all loop momenta are of the same order, and working at large $\tilde{k}$, we see that a $\ln a$ term arises from the $\tilde{k}^{2\ell-2-D\ell}$ term in the $\tfrac{1}{\tilde{k}}$ expansion of $F$ at large $\tilde{k}$, and the full $a$ dependence of this term is 
\begin{equation}
\left({1\over a}\right)^{(D-2)(\ell-1)}\ln a,
\end{equation}
which is highly suppressed in the large $a$ limit unless $\ell=1$. This verifies the above claim that only one-loop fluctuations give rise to the log term. Further, by considering different scaling regimes where various subsets of loop momenta scale to be much larger than the rest, one may verify the fact that only the two derivative sector of massless fields contributes to the log term. However, we do not do so here and instead refer the reader to \cite{Sen:2012dw} for those details.

Therefore, as argued earlier, the log term may be regarded as an IR probe of the microscopic theory, in the sense that any putative microscopic description of the black hole must correctly reproduce not only the leading Bekenstein Hawking area law, but also the log correction to it.

In this note we shall compute the log term for a class of magnetically charged extremal black holes which asymptote to AdS$_4\times S^7$ for which a complete expression for the microscopic entropy has recently been computed via the computation of a topologically twisted index in ABJM theory \cite{Benini:2015eyy}. We omit details of the microscopic formula, referring the reader to \cite{Benini:2015eyy} as well as the companion papers \cite{Benini:2015noa,Benini:2016rke}. Further, the only features of the near horizon geometry which shall be relevant to us are that it is AdS$_2\times S^2\times S^7$ where the $S^7$ is bundled over the $S^2$ and that the AdS$_2$, $S^2$ and $S^7$ have a common length scale $a$ associated to them. That is, the metric over the full 11 dimensional near horizon geometry can be brought to the form $g_{\mu\nu} = a^2 g^{(0)}_{\mu\nu}$ where the metric $g^{(0)}$ and the coordinates are $a$ independent.

Further the $S^2$ has $SO(3)$ isometry, while the $S^7$ has $U(1)^4$ isometry. These inputs suffice for the macroscopic computation of the log term as a prediction for the microscopic formula of \cite{Benini:2015eyy}. We shall finally discuss a few aspects of the proposed match. \footnote{While this draft was being readied for submission, we learned of \cite{lpz} where this comparison has been carried out in a numerical scheme to find a mismatch.} Details of the full black hole solution are reviewed in \cite{Benini:2015eyy}.
\section{The Log Term from the Quantum Entropy Function}
We will now describe how the log term may be extracted from the path integral \eqref{qef}. To do so, it is useful to phrase the problem more generally. In particular, we consider a theory in $D$ dimensions, with a dynamical field $\Phi$, admitting a saddle-point which is a background with length scale $a$. 
\begin{equation}
\mathcal{Z}\l[\Phi\r]=\int\l[\mathcal{D}\Phi\r] e^{-{1\over\hbar}S\l[\Phi\r]}.
\end{equation}
Then, as argued from Equation \eqref{scaling}, to extract the term in the free energy proportional to $\ln a$, it is sufficient to concentrate on the one-loop partition function of the theory. The techniques for this analysis are well-known and we refer the reader to \cite{bmpvlog,Bhattacharyya:2012ye,Gupta:2013sva,Gupta:2014hxa} for accounts of how these computations are carried out. 
Firstly, the one-loop partition function is then given by
\begin{equation}
\mathcal{Z}_{1-\ell}={{\det}'\mathcal{O}}^{-{1\over 2}}\cdot\left(\mathcal{Z}_{\text{zero}}\right)^{n_0},
\end{equation}
where $\det'\mathcal{O}$ is the determinant of $\mathcal{O}$ evaluated over its non--zero modes, $n_0$ is the number of zero modes of $\mathcal{O}$, and$\mathcal{Z}_{\text{zero}}$ is the residual zero-mode integral. Therefore
\begin{equation}
\ln\mathcal{Z}_{1-\ell}=-{1\over 2}\ln{\det}'\mathcal{O} + n_0\ln\mathcal{Z}_{\text{zero}},
\end{equation}
The final result is that the coefficient of the $\ln a$ term in the free energy computed about this saddle-point depends on $K\left(t;0\right)$, which is the $t^0$ coefficient of the heat kernel expansion of the kinetic operator $\mathcal{O}$ about this saddle-point, and $\beta_{\Phi}$, which determines how the zero mode contribution $\mathcal{Z}_{\text{zero}}$ to the path integral scales with $a$. 
\begin{equation}
\mathcal{Z}_{\text{zero}} = a^{\beta_{\Phi}}\hat{\mathcal{Z}}_{\text{zero}},
\end{equation}
where $\hat{\mathcal{Z}}_{\text{zero}}$ does not scale with $a$.
We eventually obtain the formula
\begin{equation}\label{logterm}
\ln\mathcal{Z}_{1-\ell}= K\left(t;0\right)\ln a+\left(\beta_{\Phi}-1\right)n_0\ln a,
\end{equation}
It is well known that in odd--dimensional spacetimes, $K\left(0;t\right)=0$ and hence we have the formula
\begin{equation}
\ln\mathcal{Z}_{1-\ell}=\left(\beta_{\Phi}-1\right)n_0\ln a.
\end{equation}
The extension to the case of multiple fields $\lbrace\,\Phi\,\rbrace$ having zero modes is apparent.
\begin{equation}\label{logtermzeromodes}
\ln\mathcal{Z}_{1-\ell}=\sum_{\phi\in\lbrace\,\Phi\,\rbrace}\left(\beta_{\phi}-1\right)n^{\phi}_0\ln a,
\end{equation}
where $n^{\phi}_0$ is the number of zero modes of the kinetic operator over the field $\phi$. The specific values for $\beta_{\Phi}$ that will be relevant to us are for the vector, the graviton, the three-form and the gravitino. These are given by
\begin{equation}\label{betas}
\beta_{v}={D-2\over 2},\quad \beta_{m}={D\over 2},\quad \beta_{f}=D-1,\quad \beta_C = {D\over 2}-3.
\end{equation}
Here $v$ denotes the vector field, $m$ the metric or the graviton, $f$ the gravitino, and the corresponding $\beta$ values
have been listed in Equation (2.37) of \cite{bmpvlog}. $C$ denotes the three-form field and its $\beta$ value may be determined exactly in the same manner as the previous fields \cite{bmpvlog,Bhattacharyya:2012ye}. We start with the expression for the normalization of the field $C_{MNP}$ in $D$-dimensions.
\begin{equation}
\int\left[\mathcal{D}C_{MNP}\right]\,\expo\left[-\int d^dx\,\sqrt{g} g^{MU}g^{NV} g^{PW} C_{MNP}C_{UVW}\right]=1.
\end{equation}
Here the metric scales as $g_{MN}= a^2 g^{(0)}_{MN}$ where $g^{(0)}$ does not scale with $a$. Therefore we have
\begin{equation}
\int\left[\mathcal{D}C_{MNP}\right]\,\expo\left[-a^{D-6}\int d^dx\,\sqrt{g^{(0)}} g^{(0)MU}g^{(0)NV} g^{(0)PW} C_{MNP}C_{UVW}\right]=1.
\end{equation}
Hence the correctly normalized integration measure is
\begin{equation}
\prod_{x,\left(MNP\right)}d\left(a^{{D\over 2}-3} C_{MNP}\right).
\end{equation}
From this we can obtain that the zero mode of $C_{MNP}$ corresponds to
\begin{equation}
\beta_{C} = {D\over 2}-3.
\end{equation}
\section{Counting the Number of Zero Modes}
From the above discussion it is clear that if we are to extract the contribution to the (logarithm of the) quantum entropy function which scales as $\ln a$, where $a$ is the length scale associated with the radii of AdS$_2$, $S^2$ and $S^7$, it is enough to compute the zero modes of the fields appearing in the path integral \eqref{qef}. Zero modes can in principle appear in the massless spectrum of AdS$_2$ fields obtained from fields of the 11 dimensional supergravity reduced on to $S^2\times S^7$. In this section we will enumerate these zero modes and compute their contribution to the quantum entropy of magnetically charged AdS$_4$ black holes. In isolating these zero modes, a special role is played by the so-called \textit{discrete modes} of the Laplacian for spin-1 and spin-2 fields, and the Dirac operator for spin-$\tfrac32$ fields on AdS$_2$ explicitly enumerated in \cite{Camporesi:1994ga,Camporesi:1995fb} respectively, and counting the total number of zero modes is essentially equivalent to counting the total number of discrete modes of these fields. These modes have also been listed in \cite{Banerjee:2011jp,bmpvlog,Gupta:2014hxa}. An important observation for us is that it turns out that naively the number of zero modes for each of these fields turns out to be infinite. However, this divergence turns out to be essentially equivalent to the the volume divergence of the free energy and may be regulated in the same way. Two slightly different, but equivalent, procedures for doing this are available in \cite{Banerjee:2010qc,Banerjee:2011jp,bmpvlog} and \cite{Gupta:2013sva,Gupta:2014hxa} and we shall use those results for the regularized number of zero modes in the computations that follow.
\subsection*{Bosonic Zero Modes}
The bosonic fields are the graviton $h_{MN}$ and the 3-form $C_{MNP}$. Quantization of the graviton gives rise to a ghost vector field but this has no zero modes. Quantization of the 3-form $C$ gives rise to a ghost 2-form $B$ with Grassmann odd statistics, and a ghost-for-ghost 1-form $A$ with Grassmann even statistics. We use the following conventions: $M$ is an 11-d vector index, $\mu$ is an AdS$_2$ index, $a$ is an $S^2$ index and $i$ is an $S^7$ index. Finally $\alpha$ is either $a$ or $i$.

Consider the metric zero modes first. The graviton $h_{MN}$ decomposes into the AdS$_2$ graviton $h_{\mu\nu}$, 3 massless AdS$_2$ vectors $h_{\mu a}$, and 4 massless AdS$_2$ vectors $h_{\mu i}$, along with the AdS$_2$ scalars $g_{ia},\,g_{ij}$ and $g_{ab}$. In counting the number of massless vectors, we used the fact that these are given by 
\begin{equation}
h_{\mu\alpha} = v_{\mu}k_{\alpha},
\end{equation}
where $k_{\alpha}$ is a Killing vector along an internal direction. Since the internal space has SU(2) $\otimes$ U(1)$^{\otimes 4}$ isometry, there are $3 + 4 =7$ Killing vectors. Each massless vector field on AdS$_2$ contributes
\begin{equation}
n_0^v = -1,
\end{equation}
hence there are $-7$ zero modes from the $h_{\mu\alpha}$. Also there are $-3$ zero modes from the AdS$_2$ metric $h_{\mu\nu}$. Therefore total number of metric zero modes is
\begin{equation}
n_0^m = -7 -3 =-10.
\end{equation}
Then the contribution to the log term from the 11d metric zero modes is 
\begin{equation}\label{logm}
\delta Z|_{\text{metric}} = \left(\beta_m-1\right) n_0^m \ln a = \left({11\over 2}-1\right)\left(-10\right) = -45\ln a.
\end{equation}
We next consider the 3-form field $C_{MNP}$. Its quantization was carried out in \cite{Siegel:1980jj,ThierryMieg:1980it,Copeland:1984qk} and is reviewed in Section 3.1 of \cite{Bhattacharyya:2012ye}. We will just require the following result. The quantization of a $p$-form requries $p$ generalized ghost fields which are $(p-j)$-forms, where $j$ runs from $1$ to $p$. Further the logarithmic contribution of all these fields to the free energy may be packaged into the expression
\begin{equation}
\Delta F = \sum_{j=0}^{p}\left(-1\right)^j\left(\beta_{p-j}-j-1\right)n^0_{\mathcal{O}_{p-j}}\ln a,
\end{equation}
where $\mathcal{O}$ is the kinetic operator. For $j=1,\ldots,p$ which are the generalized ghost fields, this is just the Hodge Laplacian. For $j=0$, which is the physical field, this can have couplings to background fluxes as well. We now consider the reduction of $C_{MNP}$ onto AdS$_2$. This firstly leads to $C_{\mu\nu a}\,C_{\mu\nu i}$ which are two-forms on AdS$_2$ which are Hodge duals of scalars. These have no zero modes. Next, we consider a 3-form which is the wedge product of a 1-form along AdS$_2$ with a harmonic 2-form along $S^2\times S^7$.
\begin{equation}
C^{(3)} = C_{(1)}^{(AdS_2)}\wedge C_{(2)}^{(S^2\times S^7)}.
\end{equation}
The number of such harmonic forms is given by the second Betti number $b_2$ of the manifold, which is 1 in this case, as may be readily seen from multiplying the Poincar\'e polynomials of $S^2$ and $S^7$. 
Hence there is a single massless vector fields along AdS$_2$ from the 11 dimensional 3-form field. Finally we have the scalars $C_{abi},\,C_{aij},\,C_{ijk}$ which don't have zero modes. Hence the 3-form contributes
\begin{equation}\label{logc}
\delta Z|_{\text{C}} = \left(\beta_C-1\right) n_0^C \ln a = \left({11\over 2}-3-1\right)\left(-1\right) = -{3\over 2}\ln a
\end{equation}
to the log term. We next consider the ghost $B_{MN}$ which arises from the quantization of $C$. This decomposes into the following massless fields on AdS$_2$. First we have $B_{\mu\nu}$ which contributes no zero modes. Next we may obtain a massless 1-form on AdS$_2$ from this field by decomposing $B_{MN}$ into a wedge product of an AdS$_2$ 1-form with a harmonic 1-form on $S^2\times S^7$. The number of such harmonic 1-forms is the first Betti number of $S^2\times S^7$ which is zero. Finally we have the scalars $B_{ab},\,B_{ai}$ and $B_{ij}$, which contribute no zero modes. Therefore the contribution to the log term is 
\begin{equation}\label{logb}
\delta Z|_{\text{B}} = -\left(\beta_B-2\right) n_0^B \ln a = 0.
\end{equation}
The overall minus sign is on account of Grassmann odd statistics of this field. Finally we have the ghost-for-ghost field $A_M$ from the quantization of $B$. This leads to one massless vector field on AdS$_2$ and therefore
\begin{equation}\label{loga}
\delta Z|_{\text{A}} = \left(\beta_A-3\right) n_0^A \ln a = \left({9\over 2}-3\right)\left(-1\right) = -{3\over  2}\ln a.
\end{equation} 
We therefore add \eqref{logm}, \eqref{logc}, \eqref{logb} and \eqref{loga} to obtain
\begin{equation}\label{logbos}
\delta Z = \left(-45-\frac{3}{2}-\frac{3}{2}\right)\ln a = -48\ln a.
\end{equation}
The residual scalar generalized ghost has no zero modes.
\subsection*{Fermionic Zero Modes}
To count fermion zero modes, we need to compute the regularized number of discrete modes $\xi^{(k)+}_\mu$ and $\hat{\xi}^{(k)+}_\mu$, $k=1,2,\ldots$, on AdS$_2$. The relevant computations are available in \cite{Banerjee:2011jp,bmpvlog,Gupta:2014hxa} and we only mention final results. Firstly, it may be shown that the regularized number of modes of both $\xi$ and $\hat{\xi}$ is given by $-1$. Further, these modes should be tensored with the spinors associated with directions transverse to AdS$_2$. This will give rise to additional multiplicity factors. To determine the multiplicity, we note first that the near horizon geometry of the black hole has a superconformal symmetry $su(1,1|1)$ with fermionic generators $G_{n}^\alpha$ where $\alpha = \pm$ and $n\in\mathbb{Z}+\tfrac12$. The gravitino zero modes we consider are associated with the generators $G_n^\alpha$ where $\vert n\vert \geq \tfrac32$. In particular, we identify $G^\alpha_{n}$ with $n\geq\tfrac32$ with $\xi^{k}$ where $n=k+\tfrac12$ and $G^\alpha_{n}$ with $n\leq -\tfrac32$ with $\xi^{k}$ where $n=-k-\tfrac12$. Hence, there is an overall multiplicity factor of 2 coming from $\alpha$ taking values $+$ and $-$. Therefore the number of fermion zero modes is 
\begin{equation}
n_0^f = (-1-1)\times 2 = -4.
\end{equation}
Then the contribution to the log term from the fermionic zero modes is 
\begin{equation}\label{logf}
\delta Z|_{\text{fermions}} = -\left(\beta_f-1\right) n_0^f \ln a = -\left(10-1\right)\left(-4\right) = +36\ln a.
\end{equation}
The overall minus is on account of Grassmann odd statistics. 
Adding \eqref{logbos} and \eqref{logf} we see that the log term is given by
\begin{equation}
\Delta F = \left(-48 +36\right)\ln a = -12\ln a.
\end{equation}
\section{On the comparison with Microscopics}
We have so far computed the log term for magnetically charged AdS$_4$ extremal black holes using the quantum entropy function. In this section we shall very briefly discuss how in principle a match with the proposed microscopic answer of \cite{Benini:2015eyy} may be carried out. At first glance one may expect that the large $N$ expansion of the logarithm of the topologically twisted index in the CFT computed in \cite{Benini:2015eyy}
may be matched term by term with the large $a$ expansion of the logarithm of the quantum entropy function. We note however, that several steps should in principle be necessary to carry out before a match can be meaningfully proposed. 

Firstly, the index computed by \cite{Benini:2015eyy} measures the black hole entropy in the grand canonical ensemble as it employs the AdS$_4$/CFT$_3$ correspondence. In contrast, the natural boundary conditions for the quantum entropy function \eqref{qef} pick out the \textit{microcanonical} ensemble \cite{Sen:2008vm}. While the choice of ensemble is irrelevant in the strict large $N$ limit, it is typically important when finite $N$ effects are taken into account. Indeed examples exist where the choice of ensemble is explicitly shown to affect the value of the log term \cite{Sen:2012dw}. Therefore, as a first step, we expect that one would need to go to the microcanonical ensemble when computing the CFT answer to match with the quantum entropy function. 

Second, it is not obvious how the naive large $N$ scalings of the log of the index are to be reproduced by the quantum entropy function. In particular, it appears from the analysis of \cite{Benini:2015eyy} that $\ln Z$ scales as
\begin{equation}
\ln Z \sim N^{3/2}+ \mathcal{O}\left(N\ln N\right).
\end{equation}
The $N^{3/2}$ term precisely matches with the Bekenstein Hawking entropy of the black hole, but the possible subleading term of order $N\ln N$ is particularly surprising from the point of view of the general scalings of contributions to the quantum entropy as expected from the quantum entropy function. In particular, on going through the scaling analysis of Equation \eqref{scaling}, it is apparent that we do not expect a term of the form $N\ln N$, and that this term should drop from the CFT$_3$ answer if a match with the quantum entropy function is to be possible.

Remarkably, it seems that a numerical estimate of the large $N$ behaviour of the index produces the pattern one would naturally expect from the quantum entropy function scalings, including an absence of the $N\,\ln N$ term, and the presence of the $\ln N$ term, albeit with a mismatching coefficient \cite{lpz}. This remarkable feature should certainly be better understood by carrying out a systematic large $N$ expansion of the CFT index while accounting for the choice of ensemble as we have indicated above. It would be interesting to return to these questions in the future.
\section*{Acknowledgements}
We would like to thank Rajesh Gupta, Nick Halmagyi, Euihun Joung, Tomoki Nosaka, Chiung Hwang, Leopoldo Pando Zayas, Ashoke Sen, Piljin Yi and Alberto Zaffaroni for helpful discussions. SL thanks the Korea Institute of Advanced Study and Kyung Hee University for hospitality while this work was carried out. SL's work is supported by a Marie Sklodowska Curie 2014 Individual Fellowship.

\end{document}